\documentclass[twocolumn,showpacs,eqsecnum,amsmath,amssymb,aps,prb,floatfix]{revtex4-1}

\usepackage{graphicx} 
\usepackage{dcolumn} 
\usepackage{bm} 
\usepackage{hyperref} 
\usepackage{slashed}
\usepackage{subfigure}
\usepackage{natbib}

\begin{document}

\title{Bilayers and quasi-3D stacks of Jain series fractional quantum Hall states from parton construction}

\author{Aditya Banerjee$^{1,2}$}
\affiliation{$^1$Harish-Chandra Research Institute, Chhatnag Road, Jhunsi, Allahabad, Uttar Pradesh 211019, India \\
$^2$Homi Bhabha National Institute, Training School Complex, Anushakti Nagar, Mumbai, Maharashtra 400085, India}

\date{\today}

\begin{abstract}
We present a description of bilayers and quasi-three dimensional stacks of Jain series of fractional quantum Hall states using their parton descriptions, and argue for them as candidate states when the interlayer coupling is comparable to the intralayer Coulomb interaction. For the bilayers, a K-matrix theory is presented and shown to be different from decoupled or bonded layers. The quasi-3D systems have gapless gauge excitations and gapped partons that may be able to move about in the 3D, thus presenting a toy model for quantum Hall-like states in three dimensions.      
\end{abstract}

\maketitle

\section{Introduction} \label{sec:introduction}

Fractional quantum Hall (FQH) effect [1] is one of the paradigmatic phenomena of strongly interacting systems in two dimensions. While the essential physics of the effect is strictly two-dimensional, much research has been done to study the consequences of the third spatial direction on FQH systems. Naturally, the main method for such purposes is to construct multilayer systems whose individual layers are FQH liquids, and then exploring the resultant phase or phases as a function of the separation between the layers (which effectively controls the interactions or tunneling of electrons between the layers).

A panoply of emergent phases can result from the coupling between the FQH layers. Couplings can come from either interlayer Coulomb interactions or interlayer electron tunneling, both of which depend on the interlayer separation $d$, and their overall effect depends on the ratio of the interlayer separation to the intrinsic magnetic length scale of the FQH layers $\ell_B \propto 1/\sqrt{B}$, where $B$ is the magnetic field. When $d/\ell_B \gg 1$, such that barely any interlayer coupling exists, the result is essentially a system of decoupled FQH layers, such as those studied in [2] for the case of Laughlin states in each layer. On the other hand, when $d/\ell_B \ll 1$ in a bilayer, the individual layers may lose their FQH identity and effectively fuse together into a new, generally non-FQH phase such as exciton superfluid states formed from pairing between particles of one layer with holes of the other [3,4], or interlayer paired composite fermion condensates [5-9] for the case when each layer is the half-filled Halperin-Lee-Read state [10]. The latter phases may also emerge when $d < \ell_B$ (but not $\ll \ell_B$), while when $d > \ell_B $ Halperin $(m,m,n)$ states [11] may emerge if interlayer tunneling is suppressed (infinite multilayered Halperin $(m,m,n)$ states have been studied in [12,13]).

However, it is much less clear, even theoretically, what the situation might be when $d \sim \ell_B$. This intermediate regime is our interest in this article. For this regime, [14] proposed a theoretical candidate multilayered state for the case of Laughlin states [15] in each layer, which we generalize to Jain series states [16] in each layer in this article. Let us first see this regime in terms of various energy scales. Each layer as a cyclotron scale $\omega = eB/m$ and an intralayer Coulomb scale $\epsilon_1 = e^2/\ell_B$. In addition, we have the interlayer Coulomb scale $\epsilon_2 = e^2/d$ and interlayer tunneling scale $t_{int}$. Firstly, we assume $\omega$ is much larger than other scales. Our regime is concerned with the situation when the interlayer Coulomb scale and interlayer tunneling scale are comparable to intralayer Coulomb scale.

The theoretical method is based on the parton description [16,17] of fractional quantum Hall states. In this description, an electron is imagined to be made up of constituent partons, which are glued together through a gluon, represented as a gauge degree of freedom (dof), which arises physically from the redundancy in the labelling of partons, or equivalently, as a Lagrange multiplier for the constraints demanding that the individual parton currents be equal to each other to be able to coherently form an electron. The deconfinement phase of the resultant parton-gauge theory corresponds to the fractionalized physics of the fractional quantum Hall states [17-19]. In [14], the authors leveraged the parton description of the Laughlin state $\nu=1/3$ to propose candidate states for the multilayered situation in the intermediate energy scale regime described above. We note that proposed partonic states may compete with a nearby Halperin states (of same filling factor) in a real experiment. 

We consider specifically two representative Jain states $\nu=2/5$ and $\nu = 2/3$, as these are experimentally the most prominent ones among the Jain heirarchy. Generalization to other Jain states is straighforward. In section II, we first review the parton construction of these states. In section III, we study bilayers of these two Jain states and analyse their effective theory in terms of the resulting K-matrix. In section IV, we study quasi-3D stacks formed from infinitely many layers of Jain states. In this scenario, a gapless gauge excitation with anisotropic dispersion emerges in the low-energy limit, and when interlayer tunneling is allowed, the partons may be able to leave their confinement to 2D layers and able to tunnel between the layers. This is the main novelty, in the quasi-3D limit, of the partonic analysis of multilayered FQH systems as introduced by [14]. In section V, we summarise our findings for the Jain states. A discussion of the effect of interlayer tunneling in the low energy theory is given in the appendix.

\section{Parton description of the Jain states} \label{sec:model1}

We shall consider the closely related cases of $\nu=2/5$ and $\nu=2/3$ FQH states. Among the states in Jain hierarchy, these two are the most prominent ones in experiments. Generalization to general states in the Jain states is straightforward. 

In the language of composite fermions (CF) [20], both the states corresponds to integral quantum Hall effect of the CFs at their effective filling of 2. However, the difference arises, within this framework, in the fact that for the $\nu=2/3$ state, the CFs see a negative (with respect to a fixed conventional direction) effective magnetic field.

We first review the so-called parton description of these states [16-17] in general terms. In this description, one imagines the electron to be made of constituent \textit{partons}, and the partons are glued together, through a gauge degree of freedom, to constitute an electron. That is, the electron operator is written as $c = f_1 f_2 f_3$. For the $\nu=2/5^{th}$ state, $f_1, f_2$ carry electrical charge of $2e/5 $ and $f_3$ carries $e/5 $. For the $\nu=2/3^{rd}$ state, $f_1, f_2$ carry electrical charge of $2e/3 $ and $f_3$ carries $-e/3 $. Clearly any relabeling of the partons is a matter of our definition and should not change the composite object which is the electron $c$ here. This means that there is a redundancy in this labeling, and this redundancy is captured through the introduction of a gauge degree of freedom (dof). Another way to say this is that, if the partons are to consistently be held together to form an electron, there must be a mediating "glue" to hold them together (in the confinement phase), and this glue should be taken into the description separately as a degree of freedom. A fractionalized emergent phase of matter in which the partons are themselves the basic degrees of freedom and not the composite object (electrons) corresponds to the deconfined phase of the resulting gauge-matter theory arising out of this description. The next step in this description is to make an \textit{ansatz} that the individual partons occupy an integral quantum Hall state themselves [16-17]. A parton description in this way is thus an effective shortcut to describe or obtain a low energy field theory for a given fractional quantum Hall state. For the $\nu=2/5^{th}$ state, $f_1, f_2$ are in $\nu'=1$ state while $f_3$ is in $\nu'=2$ state, and for the $\nu=2/3^{rd}$ state, $f_1, f_2$ are in $\nu'=1$ state while $f_3$ is in $\nu'=-2$ state. 

We note here that the more canonical ways of parton descriptions such as in [17] are based on demanding that the partons coherently form an electron, thus their individual currents be equal to each other, which results in an gauge degree of freedom coming up as essentially a Lagrange multiplier. This is not directly the case with the approach of [14] which we use, where as we will see below, a gauge degree of freedom arises from demanding that fluctuations in the hopping amplitudes of the individual partons be such that there is no fluctuation in the hopping of the composite object (the electron). As such, we see that the exact origin of the gauge degree of freedom in the two approaches is somewhat different and this is one crucial technical difference between the two approaches, and it appears to us that the latter approach is better suited in a lattice setup which is the starting point of [14] as well as ours, particularly when dealing with more than one layer of a quantum Hall system since the gauge degree of freedom in the lattice approach (arising from hopping fluctuations over a mean-field theory, as we will see below) is more amenable to the kind of multilayer scenarios studied in [14] (and in this article) than the direct field theoretical approach of [17].

\subsection{The lattice setup}
\subsubsection{Single layer case}

We begin the discussion with a lattice version of the single layer system, in which the Hamiltonian is defined on a square lattice with the (electromagnetic) flux through each plaquette taken to be $2\pi/M$ (this puts the lattice EM field periodic in a unit cell of size $M$), and the electron density to be $2/5M$ for the $\nu=2/5$ state and $2/3M$ for the $\nu=2/3$ state, with the limit $M \rightarrow \infty$. 

\begin{figure}[htp]
   
    \includegraphics[width=4cm]{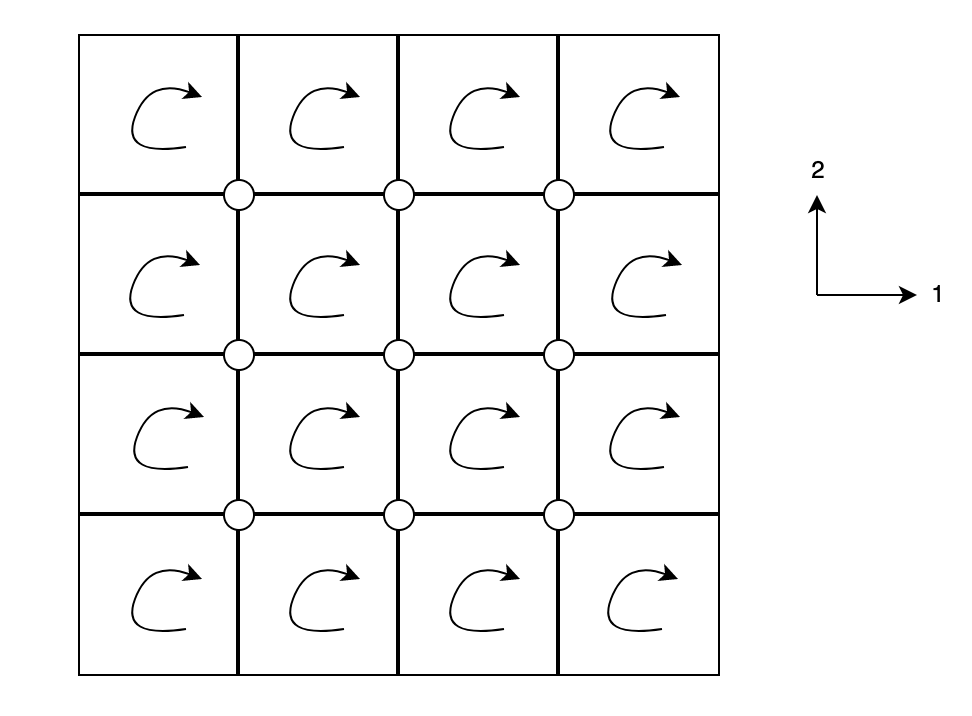}
    \caption{The square lattice with electrons (marked with circles) at the sites, and a flux (shown by the circular arrows) per plaquette of $2\pi/M$. The $1-2$ axis correspond to the $x_1$ and $x_2$ directions used in the text.}
    \label{fig:galaxy}
\end{figure}

\noindent
The lattice Hamiltonian for the electrons (denoted by creation-annihilation operators $c_{x}^{\dagger},c_x$ below) consists of the hopping terms and whichever type of interactions present between them, 

\begin{equation}
    H = -\sum_{x,i} \bigg(t c_{x}^{\dagger} e^{i\widetilde{A}_{x,i}}c_{x+\hat{x}_i} \bigg) + \textrm{interactions}
\end{equation}

\noindent
where $x$ are site indices, $i$ labels the two directions in the square lattice and the electromagnetic lattice field $\widetilde{A}_{x,i}$ is defined on the links between $x$ and $x+\hat{x}_i $ with the convention that $\widetilde{A}_{x,i}$ defines the field on the link starting at site/vertex $x$ and directed towards $i^{th}$ direction/link. Since the flux per plaquette is $2\pi/M$, we have that $\Delta_1 \widetilde{A}_{x,2} - \Delta_2 \widetilde{A}_{x,1} = 2\pi/M $, where the lattice derivative in the $\hat{x}_i$ directions is defined as $\Delta_i f_x = f_{x+\hat{x}_i} - f_x $. In the Hamiltonian above, we now substitute for $c=f_1 f_2 f_3$, which makes the hopping terms look like $t f_{3,x}^{\dagger} f_{2,x}^{\dagger} f_{1,x}^{\dagger} e^{i\widetilde{A}_{x,i}} f_{1,x+\hat{x}_i} f_{2,x+\hat{x}_i} f_{3,x+\hat{x}_i} $. At this stage the general procedure in interacting systems is to consider saddle-point/mean-field approximations which turn these three-body terms into mean-field one-body terms. Various forms of interactions may stabilize any of these saddle-point mean-field Hamiltonians, and precisely which saddle-point is stabilized depends on the details of the interactions. 

We now assume that the interactions are such that a particular type of saddle-point mean-field Hamiltonian is stabilized which leads to a parton description (in lattice form) of FQH states of our interest. This is essentially an \textit{ansatz} in any parton description of an FQH state, and different such \textit{ansatz} lead to different FQH states. We consider the mean-field Hamiltonian of the following form,

\begin{eqnarray}
    H_{mf} = && -\sum_{x,i}^{m=1,2} t_{m,x,i} f_{m,x}^{\dagger} e^{i\bar{A}^{(1)}_{x,i}}f_{m,x+\hat{x}_i} \\
    \nonumber && - \sum_{x,i}^{m=3} t_{m,x,i} f_{m,x}^{\dagger} e^{i\bar{A}^{(2)}_{x,i}}f_{m,x+\hat{x}_i} + \textrm{h.c.} 
\end{eqnarray}

\noindent
where, the lattice EM flux density seen by the parton $f_p$ ($p=1,2,3$) is $q_p$ times the EM flux density seen by the electron, where $q_p$ is the electromagnetic charge for the $p^{th}$ parton. Thus, for the $\nu=2/5$ case, the lattice EM flux densities seen by the partons are $\Delta_1 \bar{A}^{(1)}_{x,2} - \Delta_2 \bar{A}^{(1)}_{x,1} = 4\pi/(5M)$ and $\Delta_1 \bar{A}^{(2)}_{x,2} - \Delta_2 \bar{A}^{(2)}_{x,1} = 2\pi/(5M)$. Since the density of partons is the same as their parent electron, that is, $2/(5M)$, this puts $f_1$ and $f_2$ in a QH state of filling $\nu' = 1$ and $f_3$ in a QH state of filling $\nu' = 2$. Likewise, for the $\nu = 2/3$ case, we have $\Delta_1 \bar{A}^{(1)}_{x,2} - \Delta_2 \bar{A}^{(1)}_{x,1} = 4\pi/(3M)$ and $\Delta_1 \bar{A}^{(2)}_{x,2} - \Delta_2 \bar{A}^{(2)}_{x,1} = - 2\pi/(3M)$, which puts $f_1$ and $f_2$ in $\nu' = 1$ and $f_3$ in $\nu' = -2$. Here, the parton hopping amplitudes $t_{m,x,i}$ described the hopping of the parton type $m$ to the site $x$ from the site $x+\hat{x}_i$. In the mean-field situation (that is, without any fluctuations), the hopping amplitudes are independent of ${x,i}$ and can be simply written as $t_m$ (that is, constants over the lattice, but different for various partons types).

Next we consider fluctuations over this mean-field setup [23], which are considered through fluctuations in the hopping amplitudes of the form $t_{m,x,i} \rightarrow t_{m,x,i} e^{i\theta_{m,x,i}}$ with $\theta_{1,x,i} + \theta_{2,x,i} + \theta_{3,x,i} = 0$. Here, $\theta_{m,x,i}$ denote the phase of the hopping amplitude of the parton type $m$ on the link connecting $x$ and $x+\hat{x}_i$. This constraint among the $\theta_{m,x,i}$ arises because of the essential requirement that the parton hopping fluctuations should not affect the composite object's (that is, the electron's) hopping amplitude. The mean-field Hamiltonian is for the partons dynamics and likewise the fluctuations of the hopping amplitudes is for the partons, but none of this should affect the electron Hamiltonian that we began with because the electron Hamiltonian is not directly being subjected to a mean-field plus fluctuation analysis (which is being done on the partons). Hence the constraint equation for the sum of $\theta_{m,x,i}$ is that the sum should go to zero (modulo $2\pi$), which leaves the electron's hopping invariant to the fluctuations in the hopping amplitudes of the partons. Since we have three phase variables and one equation for them, we can parametrize them as $\theta_{m,x,i} = q_{mn} \mathcal{A}^{(n)}_{x,i}$, where $\mathcal{A}^{(n)}_{x,i}$ $(n=1,2)$, denote new $U(1)$ lattice gauge fields that live on the links connecting $x$ and $x+\hat{x}_i$, and we make the choice $q_{m1} = (1,-1,0) $ and $q_{m2} = (0,-1,1)$ (this choice is not unique). We thus have two $U(1)$ gauge fields to which the partons are coupled due to fluctuations. So now we have the total Hamiltonian including the fluctuations as $H=H_t + H_g$, where

\begin{eqnarray}
H_t = && -\sum_{x,i}^{m=1,2} t_{m,x,i} f_{m,x}^{\dagger} e^{i\bar{A}^{(1)}_{x,i}+iq_{mn}\mathcal{A}^{(n)}_{x,i}}f_{m,x+\hat{x}_i} \\
\nonumber && - \sum_{x,i}^{m=3} t_{m,x,i} f_{m,x}^{\dagger} e^{i\bar{A}^{(2)}_{x,i}+iq_{mn}\mathcal{A}^{(n)}_{x,i}}f_{m,x+\hat{x}_i} + \textrm{h.c.}
\end{eqnarray}

\noindent
and, the lattice gauge field dynamics term consists of the standard electric field and magnetic field terms,

\begin{equation}
    H_g = \sum_{x,i,n} \frac{g}{2} (E^{(n)}_{x,i})^{2} - \sum_{x,n} J \cos (\Delta_1 \mathcal{A}^{(n)}_{x,2} - \Delta_2 \mathcal{A}^{(n)}_{x,1})
\end{equation}

\noindent
The weak fluctuation regime of our interest is $g <<< J, t_{m,x,i} $.

\subsubsection{Multilayer case}

We can similarly give a lattice setup for the multilayered cases (assuming a general $N$ number of layers). Here, in addition to the hopping and gauge dynamics terms for each layer (that is, intralayer terms), we will have interlayer hopping and gauge dynamics terms. The physical origin of the interlayer gauge dynamics is the same as before, that is - to begin with we have an interlayer electron hopping term (but without any interlayer background EM field) which under mean-field decomposition in terms of partons yields interlayer parton hopping terms. Accounting for fluctuations of the interlayer hopping amplitudes gives rise to two $U(1)$ gauge fields which live in the direction between the layers (that is the perpendicular direction for each layer). The coupling "charges" denoted previously by $q_{mn}$ can be chosen to be the same as before, and the partons are minimally coupled to these interlayer gauge degrees of freedom with their coupling charges being $q_{mn}$.

Explicitly, in the mean-field limit we have (we add another label for the layer $z$ for the partons), 

\begin{eqnarray}
H_{mf} = \nonumber && -\sum_{x,i,z}^{m=1,2} t_{m,x,i} f_{m,x,z}^{\dagger} e^{i\bar{A}^{(1)}_{x,i}}f_{m,x+\hat{x}_i,z} \\
\nonumber && - \sum_{x,i,z}^{m=3} t_{m,x,i} f_{m,x,z}^{\dagger} e^{i\bar{A}^{(2)}_{x,i}}f_{m,x+\hat{x}_i,z}  \\
\nonumber && -\sum_{x,z}^{m=1,2} t_{m,x,3} f_{m,x,z}^{\dagger} f_{m,x,z+1} \\
&& - \sum_{x,i}^{m=3} t_{m,x,3} f_{m,x,z}^{\dagger} f_{m,x,z+1} + \textrm{h.c.} 
\end{eqnarray}

\noindent
where $t_{m,x,3}$ are the hopping amplitudes of the partons at sites labelled $x$ (in their respective layers) between the layers $z$ and $z+1$, that is, these are the interlayer hopping amplitudes. Now after considering the gauge fluctuations (which give rise to two additional $U(1)$ gauge fields living in the space between the layers), we have the total Hamiltonian,

\begin{equation}
    H = \sum_{z=1}^N (H_{t,z} + H_{g,z}) + \sum_{z=1}^{N-1} (H_{t,z,z+1} + H_{g,z,z+1})
\end{equation}
\noindent
where $H_{t,z}$ and $H_{g,z}$ describe intralayer hopping and gauge dynamics terms (the intralayer gauge field also is labelled with the layer index $z$),

\begin{eqnarray}
H_{t,z} =  && -\sum_{x,i}^{m=1,2} t_{m,x,i} f_{m,x,z}^{\dagger} e^{i\bar{A}^{(1)}_{x,i}+iq_{mn}\mathcal{A}^{(n)}_{x,i,z}}f_{m,x+\hat{x}_i,z} \\
\nonumber && - \sum_{x,i}^{m=3} t_{m,x,i} f_{m,x,z}^{\dagger} e^{i\bar{A}^{(2)}_{x,i}+iq_{mn}\mathcal{A}^{(n)}_{x,i,z}}f_{m,x+\hat{x}_i,z} + \textrm{h.c.} \\
H_{g,z} = \nonumber && \sum_{x,i,n} \frac{g}{2} (E^{(n)}_{x,i,z})^{2} - \sum_{x,n} J \cos (\Delta_1 \mathcal{A}^{(n)}_{x,2,z} - \Delta_2 \mathcal{A}^{(n)}_{x,1,z})
\end{eqnarray}

\noindent
and $H_{t,z,z+1}$ and $H_{g,z,z+1}$ describe interlayer hopping and gauge dynamics terms,

\begin{eqnarray}
H_{t,z,z+1} =  && -\sum_{x,z}^{m=1,2} t_{m,x,3} f_{m,x,z}^{\dagger} e^{iq_{mn}\mathcal{A}^{(n)}_{x,z,3}} f_{m,x,z+1} \\
\nonumber  && - \sum_{x,i}^{m=3} t_{m,x,3} f_{m,x,z}^{\dagger} e^{iq_{mn}\mathcal{A}^{(n)}_{x,z,3}} f_{m,x,z+1} + \textrm{h.c.} \\
H_{g,z,z+1} = \nonumber && \sum_{x,n} \frac{g'}{2} (E^{(n)}_{x,z,3})^{2} \\
\nonumber && - \sum_{x,i,n} J' \cos (\Delta_i \mathcal{A}^{(n)}_{x,z,3} - \mathcal{A}^{(n)}_{x,i,z} + \mathcal{A}^{(n)}_{x,i,z+1})
\end{eqnarray}
where $\mathcal{A}^{(n)}_{x,z,3}$ denotes the gauge fields corresponding to the fluctuations of the hopping amplitudes for partons at sites labelled $x$ (in their respective layers) between the layers $z$ and $z+1$. As before, our weak fluctuation regime corresponds to the coupling constants of the electric field terms being much smaller than the hopping amplitudes and magnetic field couplings.

\subsection{Single layer continuum theory} 

From here we now go to the continuum limit for the single layer case (bilayers and multilayers will be considered in the next sections). For the purpose of describing this as a continuum field theory, let us introduce for $f_1, f_2$ their respective parton gauge fields $\alpha^{(1)}$, $\alpha^{(2)}$ and write their currents as $j^{(m)}_{\mu} = \frac{1}{2\pi} \epsilon_{\mu\nu\lambda} \partial_{\nu} \alpha^{(m)}_\lambda$ for the partons $f_1 , f_2$, i.e., $m=1,2$ here. For the third parton $f_3$, since it occupies a $\nu'=\pm 2$ quantum Hall state (respectively for $\nu=2/5, 2/3$), we have to introduce two parton gauge fields $\alpha^{(3a),(3b)}$ with which to express its current as $j^{(3)}_{\mu} = \frac{1}{2\pi} \epsilon_{\mu\nu\lambda} \partial_{\nu} \alpha^{(3a)}_\lambda + \frac{1}{2\pi} \epsilon_{\mu\nu\lambda} \partial_{\nu} \alpha^{(3b)}_\lambda$. The hopping Hamiltonian above is described as an effective theory by the total Lagrangian, for the case of $\nu=2/5$ state, by the standard Chern-Simons terms for the parton fields along with the minimal coupling of their currents to the two $U(1)$ gauge fields,

\begin{eqnarray}
\mathcal{L}_{2/5} = \nonumber && \frac{1}{4\pi} \sum\limits_{m=1,2} \epsilon_{\mu\nu\lambda} \alpha^{(m)}_\mu \partial_\nu \alpha^{(m)}_\lambda + \frac{1}{4\pi} \epsilon_{\mu\nu\lambda} \alpha^{(3a)}_\mu \partial_\nu \alpha^{(3a)}_\lambda \\ 
\nonumber && + \frac{1}{4\pi} \epsilon_{\mu\nu\lambda} \alpha^{(3b)}_\mu \partial_\nu \alpha^{(3b)}_\lambda + \sum\limits_{m=1,2}^{n=1,2}\frac{1}{2\pi} \epsilon_{\mu\nu\lambda} q_{mn}\mathcal{A}^{(n)}_\mu \partial_\nu \alpha^{(m)}_\lambda \\
&& + \frac{1}{2\pi} q_{3n} \epsilon_{\mu\nu\lambda} \mathcal{A}^{(n)}_\mu \partial_\nu (\alpha^{(3a)}_\lambda + \alpha^{(3b)}_\lambda),
\end{eqnarray}
and likewise for the $\nu=2/3$ state, 

\begin{eqnarray}
\mathcal{L}_{2/3} = \nonumber && \frac{1}{4\pi} \sum\limits_{m=1,2} \epsilon_{\mu\nu\lambda} \alpha^{(m)}_\mu \partial_\nu \alpha^{(m)}_\lambda - \frac{1}{4\pi} \epsilon_{\mu\nu\lambda} \alpha^{(3a)}_\mu \partial_\nu \alpha^{(3a)}_\lambda \\ 
\nonumber && - \frac{1}{4\pi} \epsilon_{\mu\nu\lambda} \alpha^{(3b)}_\mu \partial_\nu \alpha^{(3b)}_\lambda + \sum\limits_{m=1,2}^{n=1,2}\frac{1}{2\pi} \epsilon_{\mu\nu\lambda} q_{mn}\mathcal{A}^{(n)}_\mu \partial_\nu \alpha^{(m)}_\lambda \\
&& - \frac{1}{2\pi} q_{3n} \epsilon_{\mu\nu\lambda} \mathcal{A}^{(n)}_\mu \partial_\nu (\alpha^{(3a)}_\lambda + \alpha^{(3b)}_\lambda).
\end{eqnarray}
 
\noindent
As for the continuum terms corresponding to $H_g$, we see that the electric field term is quadratic and expanding the cosine term (the magnetic field term) to leading order in the variables $\mathcal{A}^{(n)}_{x,i}$ also gives a quadratic term, so we have Maxwell terms for these gauge fields in the continuum (as expected, since the lattice terms were also Maxwell). Since integrating out the gapped partons would produce Chern-Simons terms for the dynamics of the gauge fields $\mathcal{A}_{\mu}^{(n)}$, with respect to which the Maxwell terms are irrelevant in the low-energy effective theory, we can safely ignore/drop them at the outset. We will later see that in the multilayer situations, this will generally not be the case for the interlayer gauge field that we will introduce later. For brevity we have also not explicitly written the terms corresponding to the minimal coupling of the parton currents to the external electromagnetic field.

Let us verify that our Lagrangians above do indeed describe the intended quantum Hall states of $\nu=2/5 , 2/3$. To do so, we now integrate the gauge fields $\mathcal{A}^{(n)}_{\mu}$, which produces the constraints that the parton currents are fixed to be equal to each other. That is, $j^{(1)}_\mu=j^{(2)}_\mu=j^{(3)}_\mu $. A general solution to this can be taken as $ \alpha^{(1)}_\mu=\alpha^{(2)}_\mu=\alpha^{(3a)}_\mu + \alpha^{(3b)}_\mu = \alpha_\mu $. Denoting $\alpha^{(3a)}_\mu = \beta_\mu$, we thus have two independent fields $\alpha_\mu$ and $\beta_\mu$. Substituting these in the Lagrangians above, we have,

\begin{eqnarray}
\mathcal{L}_{2/5} = \nonumber && \frac{2}{4\pi}\epsilon_{\mu\nu\lambda}\alpha_\mu \partial_\nu \alpha_\lambda + \frac{1}{4\pi}\epsilon_{\mu\nu\lambda}\beta_\mu \partial_\nu \beta_\lambda  \\
\nonumber && + \frac{1}{4\pi}\epsilon_{\mu\nu\lambda}{(\alpha-\beta)}_\mu \partial_\nu {(\alpha-\beta)}_\lambda + \frac{1}{2\pi}e A_{EM,\mu}\partial_\nu \alpha_\lambda \\
= \nonumber && \frac{3}{4\pi}\epsilon_{\mu\nu\lambda}\alpha_\mu \partial_\nu \alpha_\lambda + \frac{2}{4\pi}\epsilon_{\mu\nu\lambda}\beta_\mu \partial_\nu \beta_\lambda - \frac{1}{4\pi}\epsilon_{\mu\nu\lambda}\alpha_\mu \partial_\nu \beta_\lambda \\
&& - \frac{1}{4\pi}\epsilon_{\mu\nu\lambda}\beta_\mu \partial_\nu \alpha_\lambda + \frac{1}{2\pi}e A_{EM,\mu}\partial_\nu \alpha_\lambda 
\end{eqnarray}

\noindent
We note that we could have obtained the above Lagrangian directly in the continuum by arguing that because the partons have to be confined to each other to form an electron, their individual currents must equal each other (which is what we obtained above after integrating out the fields $\mathcal{A}^{(n)}_{\mu}$, which would have yielded the above Lagrangian directly. 

Using the K-matrix notation, we introduce a two-component vector $\Lambda = (\alpha, \beta)^T$, and $s=(1,0)$, and write the above Lagrangian as,

\begin{equation}
\mathcal{L}_{2/5} = \frac{1}{4\pi}\mathcal{K}_{2/5}\epsilon_{\mu\nu\lambda}\Lambda_\mu \partial_\nu \Lambda_\lambda + \frac{1}{2\pi}e s A_{EM,\mu}\partial_\nu \Lambda_\lambda
\end{equation}
where,

\begin{equation}
\mathcal{K}_{2/5} = \begin{pmatrix}
3 && -1 \\\\
-1 && 2
\end{pmatrix}
\end{equation}
This matrix is related to the one in [17] by a similarity transformation, which means that the theory described by our K-matrix is in the same topological class as that in [17]. In addition, the ground state degeneracy on a torus of a quantum Hall state whose effective theory is described through such a K-matrix is given by $|det(\mathcal{K})|$ [18]. We give a simple derivation of this in the Appendix. For the above K-matrix then the ground state degeneracy on torus is $5$, as it should be for the FQH state at $\nu=2/5$. 

Likewise, for the $\nu=2/3$ state, 

\begin{eqnarray}
\mathcal{L}_{2/3} = \nonumber && \frac{2}{4\pi}\epsilon_{\mu\nu\lambda}\alpha_\mu \partial_\nu \alpha_\lambda - \frac{1}{4\pi}\epsilon_{\mu\nu\lambda}\beta_\mu \partial_\nu \beta_\lambda  \\
\nonumber && - \frac{1}{4\pi}\epsilon_{\mu\nu\lambda}{(\alpha-\beta)}_\mu \partial_\nu {(\alpha-\beta)}_\lambda + \frac{1}{2\pi}e A_{EM,\mu}\partial_\nu \alpha_\lambda \\
= \nonumber && \frac{1}{4\pi}\epsilon_{\mu\nu\lambda}\alpha_\mu \partial_\nu \alpha_\lambda - \frac{2}{4\pi}\epsilon_{\mu\nu\lambda}\beta_\mu \partial_\nu \beta_\lambda - \frac{1}{4\pi}\epsilon_{\mu\nu\lambda}\alpha_\mu \partial_\nu \beta_\lambda \\
\nonumber && - \frac{1}{4\pi}\epsilon_{\mu\nu\lambda}\beta_\mu \partial_\nu \alpha_\lambda + \frac{1}{2\pi}e A_{EM,\mu}\partial_\nu \alpha_\lambda \\
= \nonumber && \frac{1}{4\pi}\mathcal{K}_{2/3}\epsilon_{\mu\nu\lambda}\Lambda_\mu \partial_\nu \Lambda_\lambda + \frac{1}{2\pi}e s A_{EM,\mu}\partial_\nu \Lambda_\lambda 
\end{eqnarray}
where,

\begin{equation}
\mathcal{K}_{2/3} = \begin{pmatrix}
1 && -1 \\\\
-1 && -2
\end{pmatrix}
\end{equation} 
As before, this matrix is also equivalent by a similarity transformation to the one in [17]. The ground state degeneracy on torus of our theory is $3$, as is should be for the FQH state at $\nu=2/3$. 

Thus we have described the effective theories of $\nu=2/5, 2/3$ starting from a parton construction. Now in the following sections, we shall use similar procedure to describe bilayers and quasi-3D limit of these states.

\section{Bilayer continuum theory} \label{sec:blyr}

\subsection{the $\nu=2/5$ case}

The case of bilayer corresponds to $N=2$ in our previous discussion of multilayer lattice setup. The main difference here from the single layer case comes from the terms corresponding to the interlayer hopping (and the consequent interlayer gauge field) of the partons. As before, the intralayer gauge field's Maxwell terms can be ignored at the outset in the low-energy limit as the more dominant Chern-Simons terms for these gauge fields will be induced by the partons. However, this is not the case with the interlayer gauge field, and thus its Maxwell terms have to be considered in the continuum theory, and this is ultimately responsible for the "coupling" between the layers in the continuum picture.

Following the conventions of [14], we enlarge, notationally, our $\mathcal{A}^{(n)}_{\mu} $ fields from having two spatial components to now having three spatial components (the first two spatial components of this enlarged notation come from the intralayer gauge fields, while the third spatial component is just the interlayer gauge field). We label the parton fields with the layer indices, $\alpha^{(m)}_{l,\mu}$, as well as the gauge fields, $\mathcal{A}^{(n)}_{l,\mu} $. The total effective action is $ \mathcal{L}_{2/5}^{(blr)} = \sum\limits_{l=1,2}\mathcal{L}_{2/5}^l + \mathcal{L}_{\bot} $ , where $i=1,2$ denotes the planar spatial indices, and,

\begin{eqnarray}
\mathcal{L}_{2/5}^l = \nonumber && \frac{1}{4\pi} \sum\limits_{m=1,2} \epsilon_{\mu\nu\lambda} \alpha^{(m)}_{l,\mu} \partial_\nu \alpha^{(m)}_{l,\lambda} + \frac{1}{4\pi} \epsilon_{\mu\nu\lambda} \alpha^{(3a)}_{l,\mu} \partial_\nu \alpha^{(3a)}_{l,\lambda} \\ 
\nonumber && + \frac{1}{4\pi} \epsilon_{\mu\nu\lambda} \alpha^{(3b)}_{l,\mu} \partial_\nu \alpha^{(3b)}_{l,\lambda} + \sum\limits_{m=1,2}^{n=1,2}\frac{1}{2\pi} \epsilon_{\mu\nu\lambda} q_{mn}\mathcal{A}^{(n)}_{l,\mu} \partial_\nu \alpha^{(m)}_{l,\lambda} \\
&& + \frac{1}{2\pi} q_{3n} \epsilon_{\mu\nu\lambda} \mathcal{A}^{(n)}_{l,\mu} \partial_\nu (\alpha^{(3a)}_{l,\lambda} + \alpha^{(3b)}_{l,\lambda}),
\end{eqnarray}
and

\begin{eqnarray}
\mathcal{L}_{\bot} = \nonumber && \sum \limits_{n=1,2} \eta_1 (\partial_{0}\mathcal{A}^{(n)}_{1,3} - \mathcal{A}^{(n)}_{1,0} + \mathcal{A}^{(n)}_{2,0})^2 \\
&& - \sum\limits_{n=1,2}^{i=1,2} \eta_2 (\partial_{i}\mathcal{A}^{(n)}_{1,3} - \mathcal{A}^{(n)}_{1,i} + \mathcal{A}^{(n)}_{2,i})^{2}.
\end{eqnarray}
where $\eta_1 = 1/g'$ and $\eta_2 = J'/2$ in terms of the coupling constants defined on the lattice previously, and the layer spacing has been put to unity.

\noindent
Let us again explain our notation here, which has been adopted from [14]. The $\mathcal{L}_{\bot}$ term is a generic Maxwell-like term for the dynamics of the interlayer gauge field (layer spacing has been put to unity here). The corresponding $U(1)$ gauge fields, which we may call $a_{l}^{(n)}$ ($l$ is the layer index, and $n=1,2$), exists only in the third (or $z$-) direction (i.e., $a_{l,3}^{(n)}$ is the only non-vanishing component of this field), while the intralayer gauge fields have spatial components in only the planar $1$-$2$ (or $x$-$y$) directions (i.e., $\mathcal{A}_{l,(1,2)}^{(n)}$). In writing the above equation, we have thus simply enlarged the notational definition (which we so far used) of $\mathcal{A}_{l,(1,2)}^{(n)}$  to contain as its third spatial component the field $a_{l,3}^{(n)}$, to write it in its enlarged form as $\mathcal{A}_{l,(1,2,3)}^{(n)}$. The $(1,2)$ components of this enlarged field now are the intralayer gauge fields while the $(3)$ component is simply the interlayer $a_{l,3}^{(n)}$. Note also that the first term in the above equation is just the electric field term in the interlayer direction and the second term is the magnetic field term in the interlayer direction. 

We now choose the gauge $\mathcal{A}^{(n)}_{1,3} = 0$, go to the basis $\mathcal{A}^{(n)}_{\pm,\mu} = \mathcal{A}^{(n)}_{1,\mu} \pm \mathcal{A}^{(n)}_{2,\mu} $, the total Lagrangian becomes,

\begin{eqnarray}
\mathcal{L}_{2/5}^{(blr)}=\nonumber && \frac{1}{4\pi} \sum\limits_{m=1,2}^{l=1,2} \epsilon_{\mu\nu\lambda} \alpha^{(m)}_{l,\mu} \partial_\nu \alpha^{(m)}_{l,\lambda} \\
  + \nonumber && \frac{1}{4\pi}\sum\limits_{l=1,2} \epsilon_{\mu\nu\lambda} \alpha^{(3a)}_{l,\mu} \partial_\nu \alpha^{(3a)}_{l,\lambda}  \frac{1}{4\pi} \sum\limits_{l=1,2}\epsilon_{\mu\nu\lambda} \alpha^{(3b)}_{l,\mu} \partial_\nu \alpha^{(3b)}_{l,\lambda} \\
+\nonumber && \sum \limits_n \eta_1 (\mathcal{A}^{(n)}_{-,0})^2 - \sum\limits_{n,i} \eta_2 (\mathcal{A}^{(n)}_{-,i})^2 \\
+\nonumber && \frac{1}{4\pi} \sum\limits_{m=1,2}^{n=1,2} \sum\limits_{\pm}\epsilon_{\mu\nu\lambda} q_{mn}(\mathcal{A}^{(n)}_{\pm,\mu} \partial_\nu (\alpha^{(m)}_{1,\lambda} \pm \alpha^{(m)}_{2,\lambda})) \\
+\nonumber && \frac{1}{4\pi} \sum\limits^{n=1,2}_{\pm} \epsilon_{\mu\nu\lambda} q_{3n}(\mathcal{A}^{(n)}_{\pm,\mu} \partial_\nu (\alpha^{(3a)}_{1,\lambda} \pm \alpha^{(3a)}_{2,\lambda} \pm \alpha^{(3b)}_{1,\lambda} \pm \alpha^{(3b)}_{2,\lambda})).
\end{eqnarray}
To obtain an effective theory in terms of the parton fields, we now integrate out the $\mathcal{A}^{(n)}_{\pm,\mu}$ fields. Clearly, integrating out the $\mathcal{A}^{(n)}_{-,\mu}$ fields generate Maxwellian terms for the parton fields which are irrelevant compared to the Chern-Simons terms and thus may be ignored hereafter. Integrating out the $\mathcal{A}^{(n)}_{+,\mu}$ fields produce the constraints that the parton currents (for each parton type) over both the layers are equal to each other, that is (suppressing the vector indices for notational clarity for now), $\alpha^{(1)}_1 + \alpha^{(1)}_2 = \alpha^{(2)}_1 + \alpha^{(2)}_2 = \alpha^{(3a)}_1 + \alpha^{(3a)}_2 + \alpha^{(3b)}_1 + \alpha^{(3b)}_2 = a$. We have eight field variables and two equations, thus there are six independent field variables. Parametrizing the various variables as $\alpha^{(1)}_1 = a-b_1$, $\alpha^{(1)}_2 = b_1 $, $\alpha^{(2)}_1 = a-b_2$, $\alpha^{(2)}_2 = b_2 $ , $ \alpha^{(3a)}_1 = b_3$, $ \alpha^{(3a)}_2 = b_4$, $ \alpha^{(3b)}_1 = b_5$, $ \alpha^{(3b)}_2 = a-b_3 - b_4 - b_5$, substituting these in the Lagrangian above and introducing the vector $\tilde{\alpha} = (a, b_1, b_2, b_3, b_4, b_5)^T$, we have the effective Lagrangian of the bilayer as,

\begin{equation}
\mathcal{L}_{bA} = \frac{1}{4\pi}\mathcal{K}_{bA}\epsilon_{\mu\nu\lambda}\tilde{\alpha}_\mu \partial_\nu \tilde{\alpha}_\lambda + \frac{1}{2\pi}e s A_{EM,\mu}\partial_\nu \tilde{\alpha}_\lambda
\end{equation}
where the charge vector is $s=(1,0,0,0,0,0)$ and 

\begin{equation}
\mathcal{K}_{bA} = \begin{pmatrix}
3 & -1 & -1 & -1 & -1 & -1 \\
-1 & 2 & 0 & 0 & 0 & 0 \\
-1 & 0 & 2 & 0 & 0 & 0 \\
-1 & 0 & 0 & 2 & 1 & 1 \\
-1 & 0 & 0 & 1 & 2 & 1 \\
-1 & 0 & 0 & 1 & 1 & 2
\end{pmatrix}
\end{equation}

Let us see some observable properties of this effective theory. The ground state degeneracy of this bilayered system on a torus is $|det(\mathcal{K}_{bA})| = 20 $. This differentiates the partonic bilayer from a system of two decoupled $\nu=2/5$ layers whose toric ground state degeneracy (by which we mean ground state degeneracy on torus) would be $25$, as from a FQH state of total filling factor $\nu=2/5 + 2/5 = 4/5$, whose toric ground state degeneracy would be $5$. 

A K-matrix theory also allows us to readily calculate the statistics of parton excitations described by the theory. To do this, let us first label the partons with their corresponding vectors $k_1 = (0,1,0,0,0,0)^T $, $k_2 = (0,0,1,0,0,0)^T $ and $k_3 = (1,0,0,0,0,0)^T $. Then, the self-exchange statistics of a parton labeled by $k_i$ is given by $\theta_i = \pi k_i^T K^{-1} k_i$ and the mutual exchange statistics between partons labeled by $k_i$ and $k_j$ is given by $\theta_{ij} = 2\pi k_i^T K^{-1} k_j$ [18,21]. 

Thus, for the K-matrix of the partonic bilayer given above, $ \mathcal{K}_{bA}$, we have $\theta_1 = \theta_2 = 7\pi/10$ and $\theta_3 = 4\pi/5$. This is distinct from the self-exchange statistical angle of $2\pi/5$ for the $2e/5$ charged excitations and $3\pi/5$ for the $e/5$ charged excitations of the $\nu=2/5$ FQH state. The mutual braiding statistics of the partons in our bilayer is given as $\theta_{12} = \theta_{23} = \theta_{13} = 4\pi/5 $.

\subsection{the $\nu=2/3$ case}

Similar considerations follow as in the case of the above subsection, except with the changes in the Lagrangians corresponding to the individual layers as described in the previous section. For the sake of not cluttering the article, we do not repeat writing the steps of the calculation, and simply present below the final expression for the effective Lagrangian of the bilayer after simplification in terms of the K-matrix, 

\begin{equation}
\mathcal{L}_{bB} = \frac{1}{4\pi}\mathcal{K}_{bB}\epsilon_{\mu\nu\lambda}\tilde{\alpha}_\mu \partial_\nu \tilde{\alpha}_\lambda + \frac{1}{2\pi}e s A_{EM,\mu}\partial_\nu \tilde{\alpha}_\lambda
\end{equation}
where the charge vector is $s=(1,0,0,0,0,0)$ and

\begin{equation}
\mathcal{K}_{bB} = \begin{pmatrix}
2 & -1 & -1 & 1 & 1 & 1 \\
-1 & 2 & 0 & 0 & 0 & 0 \\
-1 & 0 & 2 & 0 & 0 & 0 \\
1 & 0 & 0 & -2 & -1 & -1 \\
1 & 0 & 0 & -1 & -2 & -1 \\
1 & 0 & 0 & -1 & -1 & -2
\end{pmatrix}
\end{equation}

The ground state degeneracy of this bilayer system on a torus is $|det(\mathcal{K}_{bB})| = 28 $. This differentiates the partonic bilayer from a system of two decoupled $\nu=2/3$ layers whose toric ground state degeneracy would be $9$, as from a FQH state of total filling factor $\nu=2/3 + 2/3 = 1+1/3$, whose toric ground state degeneracy would be $3$. The parton self-exchange statistical angle for this case are $\theta_1 = \theta_2 = 9\pi/14$ and $\theta_3 = 4\pi/7 $. The mutual exchange statistical angles are $\theta_{12} = \theta_{23} = \theta_{13} = 4\pi/7 $.

\section{quasi-3D multilayers} \label{sec:q3d}

Let us review our procedure so far. We described the bilayers as follows - we started from a parton description of an individual layer (described by a corresponding gauge theory with Chern-Simons terms) which captures the essential low-energy physics of the layer, in addition we had an interlayer gauge degree of freedom with Maxwell-like dynamics, and finally we integrated out the various gauge fields to obtain effective K-matrix theory of the bilayered system in terms of the parton fields. This procedure can be straightforwardly generalized to constructing a quasi-three dimensional multilayer formed from stacking a very large number of layers, which we describe below.

The total Lagrangian is readily written, $ \mathcal{L}^{(3d)} = \sum\limits_{l} \mathcal{L}_{2/5}^l + \mathcal{L}_{\bot}  $, where $\mathcal{L}_{2/5}^l$ is as in eq(3.9), $l$ labels the layers, and  

\begin{eqnarray}
\mathcal{L}_{\bot} = \nonumber && \sum \limits_{n=1,2}^{l} \eta_1 (\partial_{0}\mathcal{A}^{(n)}_{l,3} - \mathcal{A}^{(n)}_{l,0} + \mathcal{A}^{(n)}_{l+1,0})^2 \\
&& - \sum\limits_{n=1,2}^{i=1,2} \sum\limits_{l} \eta_2 (\partial_{i}\mathcal{A}^{(n)}_{l,3} - \mathcal{A}^{(n)}_{l,i} + \mathcal{A}^{(n)}_{l+1,i})^{2}.
\end{eqnarray}
As before, intralayer Maxwell terms have been ignored for our purposes of being in the low-energy limit. As before, we may integrate out the intralayer and interlayer gauge fields $\mathcal{A}^{(n)}_{l,\mu}$ and generate an effective theory in terms of a large number $N$ of parton fields, and then take the limit $N\rightarrow \infty$ to access the quasi-three dimensional limit. However, this requires dealing with a very large K-matrix of dimension of the order of $N \times N$. An alternative is, for the case of many layers, to instead integrate out the parton fields and generate an effective theory in terms of the gauge fields $\mathcal{A}^{(n)}_{l,\mu}$. This, as we shall see, generates an effective theory with much smaller K-matrix, which makes the theory easier to analyse for calculating, for eg., its dispersion.

Integrating out the gapped parton fields yields the effective Lagrangian as (all summations over $n=1,2$ and $i=1,2$ below), 

\begin{eqnarray}
\mathcal{L}^{(3d)} = \nonumber && \frac{1}{4\pi}\sum\limits_l \epsilon_{\mu\nu\lambda}\mathcal{K}\mathcal{A}_{l,\mu}^{T} \partial_\nu \mathcal{A}_{l,\lambda} \\
\nonumber && + \sum \limits_{n,l} \eta_1 (\partial_{0}\mathcal{A}^{(n)}_{l,3} - \mathcal{A}^{(n)}_{l,0} + \mathcal{A}^{(n)}_{l+1,0})^2 \\
&& - \sum\limits_{n,i,l} \eta_2 (\partial_{i}\mathcal{A}^{(n)}_{l,3} - \mathcal{A}^{(n)}_{l,i} + \mathcal{A}^{(n)}_{l+1,i})^{2},
\end{eqnarray}
where we defined for the intralayer components in a given layer $\mathcal{A}_{l,\mu} = (\mathcal{A}^{(1)}_{l,\mu},\mathcal{A}^{(2)}_{l,\mu})^T$ and,

\begin{equation}
\mathcal{K} = \begin{pmatrix}
-2 && -1 \\
-1 && -3
\end{pmatrix}
\end{equation}

It is readily seen that for the purpose of deriving the effective theory in terms of $\mathcal{A}_\mu$ fields, the system formed from many layers of $\nu=2/3$ states has the exact same form of the effective theory except with the resultant $\mathcal{K}$-matrix for $\mathcal{A}_{\mu}$ is simply the negative of the above equation, which reveals that as far as universal properties are concerned, both the stacked systems approach the same quasi-3D system.

From eq (3.14), we can readily find the dispersion of the gauge field. The Fourier transformed Lagrangian (note that Fourier transformation does away with the layer indices), in the basis $\mathcal{A}_{\pm,\mu} = \frac{1}{\sqrt{2}}(\mathcal{A}^{(1)}_{\mu} \pm \mathcal{A}^{(2)}_{\mu})$ with temporal gauge $\mathcal{A}_{\pm,0} = 0$, can be written as $ \mathcal{L}^{(3d)} (\vec{p}) = \sum\limits_{r=\pm}{\mathcal{A}_r (-\vec{p})}^{\dagger} L_r (\vec{p}) \mathcal{A}_r (\vec{p})$, where

\begin{equation}
L_r = \begin{bmatrix}
-\eta_2 p_z^2 &&& \frac{ik_r p_o}{4\pi} &&& \eta_2 p_1 p_3 \\\\
-\frac{ik_r p_o}{4\pi} &&& -\eta_2 p_z^2 &&& \eta_2 p_2 p_3 \\\\
\eta_2 p_1 p_3 &&& \eta_2 p_2 p_3 &&& \eta_1 p_0^2 - \eta_2 (p_1^2 + p_2^2)  
\end{bmatrix}
\end{equation}
where $k_r = -1.382, -3.618$ are eigenvalues of $\mathcal{K}$. From $L_r$, we find a gapless mode with dispersion,

\begin{equation}
\epsilon_r^{2} = \frac{\eta_2}{\eta_1}(p_1^2 + p_2^2) + \frac{16 \pi^2 \eta_2^2}{k_r^2} p_3^4
\end{equation}
Thus, for the realistic case of large but finite $N$ number of layers, $p_3 \sim \pi/N$, and the lowest mode has the dispersion,

\begin{equation}
\epsilon_r = \frac{\eta_2}{\eta_1}(p_1^2 + p_2^2) + \frac{16 \pi^6 \eta_2^2}{k_r^2 N^4}.
\end{equation}

Similar to the Laughlin case, the gauge mode disperses anisotropically with similar form of the dispersion relation as in the Laughlin case of [14]. The quasi-3D system has gapless (or nearly gapless, in a realistic situation) gauge boson excitation, as well as gapped partons $f_i$, which are minimally coupled to it and therefore have long-range interactions, and moreover, their interactions are naturally anisotropic. In addition, the partons due to being minimally coupled to the gauge field component in the $z-$direction are thus free to move around in the 3D bulk and not confined to the 2D layers of their original FQH states. However, the parton excitations are only fractionally charged by construction but can not have fractional statistics in this quasi-3D system since point-particle-like excitations can not have fractional statistics in 3D. 

Let us also comment on boundary transport of the quasi-3D system. The boundary transport is now a sheet-like surface transport, with components in the $xy-$direction, as well as in the $z-$direction, the former being quantized due to arising from the QH nature of the individual layers in the bulk, while the latter is unquantized and non-topological. In the large but finite $N$ layers, there are order $N \times N$ chiral modes of quantized transport in the $xy-$planes due to the bulk K-matrix (in the effective theory of the partons) being order $N \times N$. In the 3d limit, however, due to the gapless gauge mode in the bulk, the quantized $xy$-transport may continue to be gapless and chiral due to the underlying QH nature of the layers in the bulk, however, the gapless gauge mode decoheres the $z$-transport, which is non-quantized to begin with and is basically an artifact of the interlayer hopping events. The gapless mode in the bulk may also affect the \textit{a priori} chiral $xy-$transport due to the lack of energetic distinction between these gapless edge modes and the bulk gapless mode. This eventual lack of distinction may serve as an experimental signature of the existence of a bulk gapless mode in such quasi-3D multilayered systems, in particular, by closely observing such vanishing energetic distinction between the gapless chiral edge modes and gapless bulk gauge mode as a function of the number of layers.

\section{Summary} 
 
In this article, we have extended the multilayered Laughlin partonic theory of [14] to the case of bilayered and multilayered quasi-3D stacks of Jain series FQH states using their parton descriptions, suggesting that these may be useful candidate states for the situation when the interlayer coupling (via interlayer hopping) is comparable to the intralayer interactions. For the bilayered cases, we presented a K-matrix theory from which we showed how the bilayers represent experimentally different QH states from the cases when the layers are either decoupled or have bonded together to form an additive QH liquid. For the quasi-3D stacks, we constructed its effective theory in terms of the gauge fields which have a gapless anisotropically dispersing mode in the low-energy, and discussed its plausible consequences on the observable edge/surface transport. The most important future direction would be to construct ways to analyse the surface sheet transport, which, as we mentioned earlier, has not been successful even for the simpler case of $\nu=1/3$ FQH states in each layer.

\section{Acknowledgements}

I would like to thank Prof. Sumathi Rao for discussions and comments on improving the manuscript. Financial support from Infosys Foundation through "Senior Students Prize" and Department of Atomic Energy, India are acknowledged.

\section{References}

[1] Perspectives in Quantum Hall Effects, Eds. S. Das Sarma, A. Pinczuk, Wiley Interscience, (2007).

[2] L. Balents and M. P. A. Fisher, Phys. Rev. Lett. {\bf 76}, 2782 (1996).

[3] J. P. Eisenstein and A. H. MacDonald, Nature {\bf 432}, 691 (2004).

[4] J. P. Eisenstein, Annu. Rev. Cond. Matt. Phys. {\bf 5}, 159 (2014).

[5] N. E. Bonesteel, I. A. McDonald and C. Nayak, Phys. Rev. Lett. {\bf 77}, 3009 (1996).

[6] T. Morinari, Phys. Rev. B. {\bf 59}, 7320 (1999).

[7] G. Moller, S. H. Simon and E. H. Rezayi, Phys. Rev. Lett. {\bf 101}, 176803 (2008).

[8] J. Alicea, O. I. Motrunich, G. Rafael and M. P. A. Fisher, Phys. Rev. Lett. {\bf 103}, 256403 (2009).

[9] H. Isobe and L. Fu, Phys. Rev. Lett. {\bf 118}, 166401 (2017).

[10] B. I. Halperin, P. A. Lee, and N. Read, Phys. Rev. B. {\bf 47}, 7312 (1993).

[11] B. I. Halperin, Helv. Phys. Acta {\bf 56}, 75 (1983).

[12] J. D. Naud, L. P. Pryadko and S. L. Sondhi, Phys. Rev. Lett. {\bf 85}, 5408 (2000).

[13] J. D. Naud, L. P. Pryadko and S. L. Sondhi, Nucl. Phys. B. {\bf 594}, 713 (2001).

[14] M. Levin and M. P. A. Fisher, Phys. Rev. B. {\bf 79}, 235315 (2009).

[15] R. B. Laughlin, Phys. Rev. Lett. {\bf 50}, 1395 (1983).

[16] J. K. Jain, Phys. Rev. B {\bf 40}, 8079 (1989).

[17] B. Blok and X.-G. Wen, Phys. Rev. B {\bf 42}, 8133 (1990).

[18] X.-G. Wen and A. Zee, Phys. Rev. B. {\bf 58}, 15717 (1998).

[19] E. Fradkin, \textit{Field Theories of Condensed Matter Physics}, Cambridge University Press (2013).

[20] J. K. Jain, \textit{Composite Fermions}, Cambridge University Press (2007).

[21] X.-G. Wen and A. Zee, Phys. Rev. B {\bf 46}, 2290 (1992).

[22] X.-G. Wen and A. Zee, Phys. Rev. B {\bf 47}, 2290 (1993).

[23] We thank M. Levin for a helpful correspondence regarding hopping fluctuations.

\section{Appendix - Toric ground state degeneracy}

In this appendix, we will present a brief derivation of the ground state degeneracy on a torus of a quantum Hall fluid in the K-matrix description, following [18]. 

Let us first show that the toric degeneracy of a Chern-Simons theory at level $k$ is $|k|$. This corresponds to a $1 \times 1$ $\mathcal{K}$-matrix, i.e., $\mathcal{K} = k$, and would describe the effective theory of a fractional quantum Hall liquid at filling $\nu=1/k$. That is, we have 

\begin{equation}
\mathcal{L} = \frac{k}{4\pi}\epsilon_{\mu\nu\lambda}\alpha_{\mu}\partial_{\nu}\alpha_{\lambda}.
\end{equation}

Consider a torus of dimensions $(L_1,L_2)$. We may define gauge field configurations (essentially a gauge choice) on the torus as 

\begin{equation}
\nonumber  \alpha_0 (x,y,t) = 0, \\
\alpha_1(x,y,t) = \frac{2\pi X}{L_1}, \\
\nonumber  \alpha_2(x,y,t) = \frac{2\pi Y}{L_2},
\end{equation}
where $X$ and $Y$ are periodic coordinates on the two big circles of the torus.

Substituting this configuration into the Chern-Simons Lagrangian above gives us,

\begin{equation}
\mathcal{L} = \pi k ( Y\dot{X} - X \dot{Y}).
\end{equation}
On the torus, there exist two large gauge transformations, $g_1 = \exp(2\pi i x/L_1)$ and $g_2 = \exp(2\pi i y/L_2)$, under which $\alpha_\mu \rightarrow \alpha_\mu - ig_{1,2}^{-1}\partial_\mu g_{1,2}$. This is readily seen to transform $(X,Y) \rightarrow (X+1,Y)$ and $(X,Y) \rightarrow (X,Y+1)$ for $g_1, g_2$ respectively. Thus $(X,Y) \sim (X,Y+1) \sim (X+1,Y) $ is an equivalence relation.

From the substituted Lagrangian above, we have the "momentum" conjugate to $Y$, 

\begin{equation}
p = \frac{\delta \mathcal{L}}{\delta \dot{Y}} = -2\pi k X,
\end{equation}
so that we have the noncommutativity relation,

\begin{equation}
\left[X,Y \right] = \frac{-i}{2\pi k},
\end{equation}
and the Hamiltonian vanishes, a defining feature of a "pure" topological field theory,

\begin{equation}
\mathcal{H} = p\dot{Y} - \mathcal{L} = 0.
\end{equation}

Thus it appears \textit{a priori} that any arbitrary function qualifies as the ground-state wavefunction of such a vanishing Hamiltonian. However, legitimate wavefunctions need to satisfy the aforementioned equivalence relation. This condition filters out a set of finite number of ground state wavefunctions, and the number of elements of this set is the ground state degeneracy of the level-$k$ Chern-Simons theory on the torus.

The condition $\psi(Y) = \psi(Y+1)$ implies we can write, for integer $n$,

\begin{equation}
\psi(Y) = \sum_{n=-\infty}^{\infty} c_n \exp(2i\pi nY).
\end{equation}
Since $X$ is essentially a conjugate momentum variable to $Y$, we have to Fourier transform the above wavefunction to impose the condition $\psi(X) = \psi(X+1)$, with $p=i\partial/{\partial Y}$,

\begin{equation}
\tilde{\psi}(p) = \sum c_n \delta(p-2\pi n).
\end{equation}  
Since, as noted above, $p=-2\pi k X$, we may equivalently write, 

\begin{equation}
\phi(X) = \sum c_n \delta (kX+n).
\end{equation}
$X \sim X+1$ thus implies that $c_n = c_{n+k} $, and thereby that there are $k$ independent $c_n$'s. This implies that the number of degenerate ground state wavefunctions on the torus is $k$. 

Same conclusion is reached if we had instead chosen $X$ as our canonical position variable and $Y$ as its corresponding conjugate momentum.

We recall that the above derivation was done for a $1 \times 1$ $\mathcal{K}$-matrix. Now consider a $ m\times m$ $\mathcal{K}$-matrix which we assume to be invertible so that there is no eigenvalue $=0$, and which in general may have non-zero off-diagonal entries. There always exists a diagonalizing transformation to a new basis (new gauge fields $\alpha^{'}$) which brings the $\mathcal{K}$-matrix to its diagonal form with its eigenvalues being the diagonal entries, and thus the determinant (ground state degeneracy on torus) is simply the product of the eigenvalues. Since eigenvalues, and thus determinant, of a matrix are independent of basis or similarity transformations, it follows that the ground state degeneracy on torus of a general invertible $m \times m$ $\mathcal{K}$-matrix Chern-Simons theory is simply $|det(\mathcal{K})|$.

\end{document}